# Enhancing Agile Software Development Sustainability through the Integration of User Experience and Gamification

Manal Alhammad[1][0000-0001-7789-6667] and Ana Moreno[2][0000-0002-7815-0346]

[1] King Saud University, Riyadh, Saudi Arabia
`manalhammad@ksu.edu.sa`
[2] Universidad Politécnica de Madrid, Madrid, Spain
`ammoreno@fi.upm.es`

**Abstract.** This article provides a rich discussion on how the sustainability of agile development processes can be enhanced. In particular, we focus on a recently developed framework, named GLUX, that integrates Lean UX into Scrum. GLUX's main goal is to facilitate a seamless integration between agile and user experience (UX) by using gamification to motivate agile teams to adopt a user-centered mindset and carry out UX activities collaboratively throughout the development process. Our role as software researchers is to contribute towards improving software sustainability and provide the software engineering community with the tools and techniques that will improve the human, economic, and environmental sustainability of software development. We found that GLUX addresses human sustainability by empowering self-sufficient, problem-focused teams, building a motivating and engaging environment, and developing team cooperation. Economic sustainability is addressed by minimizing UX debt and using gamification techniques to direct the focus of the behavior and mindset of agile teams towards value creation. Finally, environmental sustainability is promoted by encouraging agile teams to build a minimum viable product (MVP).

## 1   Introduction

There is no question nowadays that software has a major impact on sustainability [1]. As Calero et al. in [2] state, software can be considered as part of the problem, but also has a lot to do with the solution. These same authors present a literature review of how sustainability is addressed in software development [3]. They identify two different perspectives. On one side, they look at what is referred to as software sustainability (SOS), which is concerned with how to make software production more sustainable, covering everything from the people, through the process and the business, to the product. On the other hand, some literature considers software as part of sustainability (SAPOS), where software is considered as a new dimension of this attribute.

Sustainable software engineering (SSE) is an emerging discipline that aims to address the long-term impact of designing, building, deploying, and maintaining software products [4]. In this vein, Calero and Piattini identify three dimensions of software sustainability, in line with the standard definition of sustainability, based on the three types of resources (human, economic, energy) essential in the software life cycle processes [3]. Human sustainability refers to analyzing and addressing the impact of software development on the sociological and psychological aspects of the software engineering community and its individuals. Economic sustainability refers to the means by which the software process protects stakeholder investments, ensures benefits, reduces risks, and maintains assets. Environmental sustainability refers to the impact of software development and maintenance on energy consumption and the usage of other resources.

Sustainability has also been recently addressed in Agile software development. For example, in [5], Eckstein and Melo analyze how sustainability is promoted by the agile philosophy, and give a detailed discussion about how it is considered in each of the agile principles. Obradović, Todorović, and Bushuyev in [6] discuss the relation between agile project management and sustainability and conclude that they are overlapping and that agile project management requires the implementation of sustainability aspects. Ochoa-Zambrano in [7] examines how collective intelligence fits into the agile manifesto and its values and principles and discusses the role of collective intelligence as a tool to enhance agile and sustainability by emphasizing team collaboration and learning. Additionally, there are a few attempts to modify the agile development process to incorporate practices that contribute to making more sustainable software products [8].

In this paper, we look into how user experience (UX) and gamification can reinforce the human, economic, and environmental sustainability of the agile development process. Through the lens of SOS, we particularly examine the recently developed GLUX (Gamified Lean UX) framework whose main goal is to facilitate a seamless integration between agile and UX by using gamification to motivate agile teams to adopt a user-centered mindset and carry out UX activities collaboratively throughout the development process.

## 2  What is GLUX?

Endeavors to combine agile and UX are based on the premise that both disciplines share common principles, such as iterative development, emphasis on the user, and team coherence [9]. Even though the potential benefits of this integration are recognized, existing approaches have been found to have a number of limitations [10]. For example, the process is not fully integrated [11], agile developers do not have a UX mindset [12], or there is a lack of rigorous empirical studies [9].

In order to address some of the challenges of integrating agile and UX, we have developed the GLUX framework [13]. GLUX engages practitioners to collaboratively integrate Lean UX activities within Scrum. Lean UX is a lightweight and iterative UX design process that is based on design thinking, lean startup and Agile [14]. GLUX takes into account the different characteristics and complexity of software engineer and agile team motivational factors. It is aimed first and foremost for small Scrum teams who are struggling with integrating UX activities into the development process, particularly in the absence of UX specialists. Even if a UX specialist is on hand, the GLUX framework can help align the workflow of the whole team towards building user-centered software projects through a shared understanding. As illustrated in **Fig. 1**, the GLUX framework is divided into two fundamental parts:

- Five Lean UX tactics for integration into Scrum: Hypotheses, Design Studio, Experiment Stories, Minimum Viable Product (MVP), and Weekly User Experiments.
- A customizable gamification strategy based on three game techniques: rewards (points and badges), challenges, and levels.

The full documentation of the GLUX framework is available at https://bit.ly/2Xp1H1A.

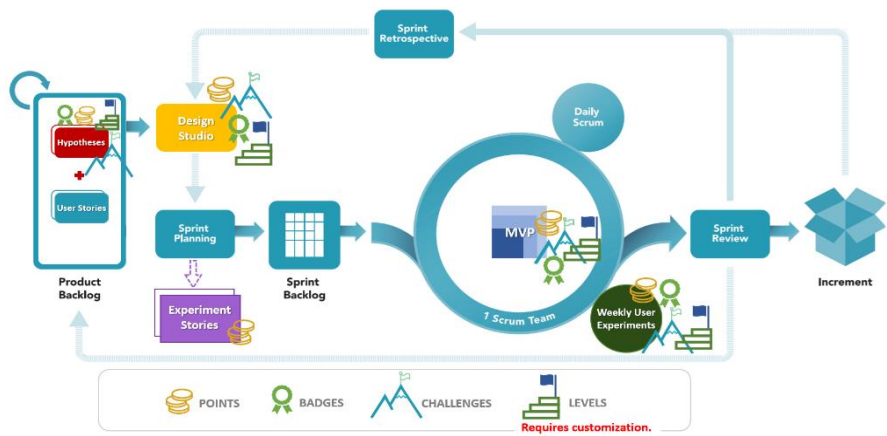

**Fig. 1.** A general overview of the GLUX framework, integrating Scrum with Lean UX using gamification.

A typical scenario of an Agile team following GLUX is as follows: the team creates a list of hypotheses on their assumptions about the needs of potential users, which they discuss with the product owner (PO) during product backlog refinement. The sprint kicks off with the design studio, where the team picks a hypothesis as a theme to guide the work of the sprint and start sketching and discussing design ideas accordingly. The sprint planning meeting should take place immediately after the design studio. Apart from deciding the user stories to be developed, the team also plans for the weekly user experiment during sprint planning, in which the team puts a version of their developing product, i.e., an MVP, to the test in order to gather feedback from the user. User experiment plans are captured in what is called an experiment story.

The team is rewarded for applying each Lean UX tactic and receives additional rewards for doing it collaboratively. The Scrum team is also encouraged by special rewards for employing Lean UX tactics for the first time, as well as for addressing some Lean UX challenges. Challenges are typically set by the team based on current UX-related difficulties and issues.

## 3 How does GLUX address the three dimensions of SOS?

GLUX was applied for two academic semesters into a novel graduate software engineering course offered as part of the MS in Software Engineering program at the Universidad Politécnica de Madrid. The discussion below is based on the lessons learned and the preliminary evidence that we got from this experience reported in [19].

### 3.1 Human sustainability

Human sustainability deals with sociological and psychological aspects of software development and developers [3]. Such aspects are critical in an intellectual activity like software construction [15].

In this sense, as discussed in the previous section, one of the main challenges to Agile UX is that developers do not have a UX mindset [13]. This challenge means that agile teams continuously prioritize delivering fast and "working" features, while paying less attention to UX aspects. This tendency mostly emerges from four key issues. First, the Agile Manifesto term "working software" came over time to be misinterpreted, and agile teams began to shift their focus to delivering functioning software rather than to supplying valuable software. Second, due to the lack of knowledge on and training in UX design, software engineers have very little motivation for and interest in UX work. Third, it is argued that "resistance to change" is linked to the reluctance to integrate UX practices into agile development. The last issue is related to what is called the curse of knowledge in which agile teams may come to believe that they know better than the user what features to build and how they should be designed and delivered.

In GLUX, a UX mindset is explicitly promoted and rewarded in four ways:

**Empowering teams to become self-sufficient.** GLUX aims to empower agile teams with the skills and mindset needed to facilitate the integration of UX into agile through a cohesive process (**Fig. 1**) vs. a parallel or dual track. This would ultimately help the team to consider the Agile-UX process as a single collaborative process in which UX work is part of the agile development process.

**Establishing a problem-focused team.** Rather than providing a set of features for implementation, GLUX guides agile teams to strive for continual improvement and builds trust within teams which take part in designing the solutions that can help to address a business outcome or a user need in the form of hypotheses, design sketches, or MVPs. This can give teams a greater sense of pride, control, and ownership over the solutions they came up with.

**Building a motivating and engaging environment.** As discussed earlier, Agile UX faces to different challenges. Darin et al. in [16] found that engaging the development team in the user-centered design (UCD) process would help make the integration a more natural process. They also emphasized the importance of keeping the team motivated and committed. Some agile teams need extrinsic motivators to perform UX activities. The GLUX framework aims to support agile teams in being more proactive about UX work through its gamification strategy. Gamification in SE is seen as a promising technique to improve software engineers' motivation and engagement and to pro-mote SE best practices [17]. The gamification strategy was designed based on a comprehensive analysis of the nature and characteristics of agile teams [13]. For instance, we found that one of the key motivators of agile teams is having a sense of achievement and a regular stream of feedback on the work they do [18]. As a result, GLUX's gamification strategy includes a rewards system, a challenges system, and a levels system. The rewards system provides the team with a recognition of their achievements. The challenges system provides the team with a sense of achievement and encourages the entire team to be more goal-driven, focused, and collaborative. The levels system provides the team with a sense of progression, which can consequently improve the team's motivation to collaborate and apply Lean UX tactics. Furthermore, GLUX's gamification strategy establishes a fun environment t using elements such as badges and an achievements board. **Fig. 2** shows also a few examples of the cheat cards we created for the master's course[1] where we applied GLUX [19]. Cheat cards are essentially a visual summary of each Lean UX tactic in terms of the rules for integrating such tactics in Scrum, how the scoring system works, and one proposed challenge.

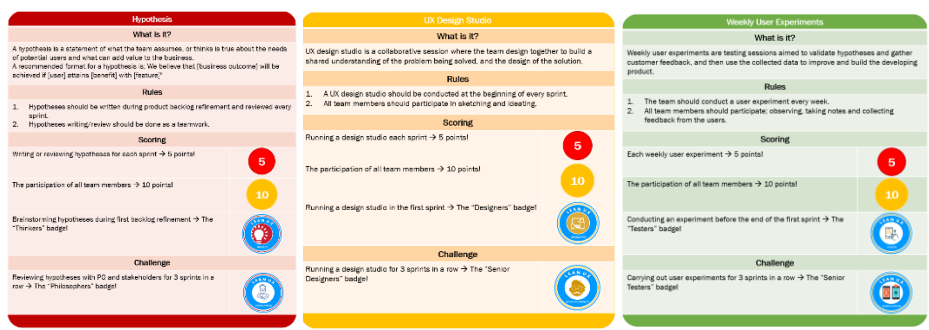

**Fig. 2.** Three examples of GLUX Cheat Cards, providing a visual summary of GLUX's rules and scoring system.

**Developing a cooperative team.** GLUX's gamification strategy is team based, where the whole team is encouraged to participate in some activities that require collaboration for the team to earn the associated rewards. Rewarding the collaboration of the entire team is aimed at facilitating a frequent and quick way to exchange ideas, allowing the team to move forward quickly into the right direction with everyone having a clear picture of the envisioned increment in each sprint. In addition, collaboration among team members from different backgrounds and expertise can help the team to learn from each other (for example, during the design studio developers can understand some aspects of the proposed design or suggest some changes from their perspective, and designers can understand the technical complexity of implementing a particular design). Ultimately, collaboration, experience sharing, and learning from each other can enhance the work performed by individuals in the future and encourage them to pass on the knowledge and skills they learned to other teams.

---

[1] https://bit.ly/GLUXGUIDE2

### 3.2 Economic sustainability

Economic sustainability is either explicitly or implicitly promoted in some of the 12 agile development principles [5]. For example, the eighth principle, which reads "Agile processes promote sustainable development. The sponsors, developers, and users should be able to maintain a constant pace indefinitely", promotes economic sustainability explicitly by advocating the development of valuable features and avoiding what is known as feature creep. On the other hand, agile development promotes economic sustainability implicitly by enabling the team to measure the economic impact of the delivered product and constantly learn about and improve both the product and the process.

GLUX addresses economic sustainability by **minimizing UX debt** to reduce rework. If UX issues are not given as much attention as functional issues, it will be all too easy for UX debt to creep in and pile up. UX debt result in teams building up UX deficiencies and then being forced to carry out additional rework. This phenomenon would eventually have a bearing on key performance indicators, as it would also impact developer productivity. More UX debt means extra hours of work at greater expense for the business. While a few other factors may lead to additional working hours, the causes can usually be traced back to the failure to focus on continuous quality improvement [20]. In GLUX, UX debt is addressed through a gamification technique called sprint challenge. The sprint challenge encourages prompt and collaborative proactivity with respect to UX debt issues. It starts with the team identifying the challenging process issues related to UX that they face and which they have listed from easiest to hardest. Starting with the easiest, the team picks a challenge every other sprint, sets it as the sprint challenge and works towards addressing this question. A sprint challenge keeps the team more focused on the pending UX issues that may turn into UX debt and result in rework. This way, the team is able to respond rapidly and at a much lower cost. It also helps teams to be more proactive about new technologies or new opportunities in any form.

Economic sustainability is also reinforced in GLUX mainly through the **hypothesis definition**. By creating a hypothesis, agile teams map every feature they intend to develop to a business outcome and a user benefit. Any feature or task that does not contribute to achieving or improving a business outcome or a user benefit is considered unnecessary. The more unnecessary features the team can avoid, the less resources are wasted. Additionally, Lean UX offers a canvas to prioritize hypotheses based on risk and value [21]. Risk refers to how damaging it would be for the business or product if the team was wrong about this hypothesis. Value refers to the perceived value that will be generated from developing the feature. Based on this prioritization, the team should only spend their resources on testing hypotheses with high risk and high value.

### 3.3 Environmental Sustainability

According to Calero et al. in [22], the key to software sustainability is to improve its power consumption. However, software, unlike hardware, production and usage has not witnessed continuous advancements in terms of energy efficiency [23]. GLUX addresses environmental sustainability in two ways. First, by **building just enough product** to validate the hypothesis. Before jumping into building a complete feature or increment, agile teams are encouraged to build an MVP, that is, the simplest version of the envisioned feature. The MVP is then validated with real users to check whether the feature provides value to the user or the business. If the feature is found to be useful, the team works on improving that feature and releases it to the end user. If not, the team can roll back that feature with a minimal environmental cost. Building a hypothesis-driven MVP can minimize energy waste by building the smallest piece of software to validate the usefulness of that feature. Second, GLUX addresses environmental sustainability by promoting collaboration in-between developers, designers, product owners, and project managers on most activities. With increased **cross-functional and continuous collaboration** among the whole team, fewer resources are used for documentation and correspondence.

## 4   Conclusion

In this paper, we examined how the integration of Lean UX and gamification can enhance the sustainability of agile development processes from the human, economic, and environmental perspectives. We presented a recently developed framework called GLUX. Through gamification, GLUX aims to engage agile teams in integrating Lean UX tactics into Scrum in a collaborative way. We have seen how human sustainability can be addressed in GLUX by empowering self-sufficient teams, establishing problem-focused teams, building a motivating and engaging environment, and developing a cooperative team. We have also discussed how GLUX addresses economic sustainability by minimizing UX debt and using gamification techniques to influence the behavior and mindset of agile teams to focus on creating value and reducing waste. Finally, we looked into how GLUX promotes environmental sustainability by encouraging agile teams to build just enough product to validate their hypothesis and continuously collaborating on most activities throughout the development process.

We acknowledge that there are not enough efficient and validated techniques and tools to enhance software sustainability. We have used GLUX in an academic setting with promising results [19]. However, it is fundamental to empirically validate how GLUX can contribute to improving the sustainability of software development in a real-world context. Partnering with agile teams from industry would be useful to verify the feasibility of the proposed ideas.

It is also important in this context to discuss how highly usable software products might have the opposite effect on sustainability. Simple and effective software might in fact encourage people to use it more often and get more people to use it as well without a specific purpose. This paradox is related to the so-called rebound effect, which is generally defined as "the difference between the expected and the actual environmental savings from efficiency improvements" [24]. Promoting the responsible consumption of software products can lead to a win-win situation, where users are more aware and engaged as active participants in sustainability practices.